\documentclass[12pt]{iopart}

\usepackage{iopams}
\usepackage{graphicx}

\newcommand {\barr} {\begin{eqnarray}}
\newcommand {\earr} {\end{eqnarray}}
\newcommand {\beq} {\begin{equation}}
\newcommand {\eeq} {\end{equation}}

\begin{document}

\title[Node or not in LaFePO, BaFe$_2$(AsP)$_2$, and KFe$_2$As$_2$ ?]{Are there nodes in LaFePO, BaFe$_2$(AsP)$_2$, and KFe$_2$As$_2$ ?}

\author{Yunkyu Bang}

\address{Department of Physics,
Chonnam National University, Kwangju 500-757, Republic of Korea}
\ead{ykbang@chonnam.ac.kr}
\begin{abstract}
We reexamined the experimental evidences for the possible
existence of the superconducting (SC) gap nodes in the three most
suspected Fe-pnictide SC compounds: LaFePO,
BaFe$_2$(As$_{0.67}$P$_{0.33}$)$_2$, and KFe$_2$As$_2$. We showed
that while the $T$-linear temperature dependence of the
penetration depth $\lambda(T)$ of these three compounds indicate
extremely clean nodal gap superconductors, the thermal
conductivity data $\lim_{T,H \rightarrow 0} \kappa_S (H,T)/T$
unambiguously showed that LaFePO and
BaFe$_2$(As$_{0.67}$P$_{0.33}$)$_2$ are extremely dirty, while
KFe$_2$As$_2$ can possibly be clean. This apparently conflicting
experimental data casts a serious doubt on the nodal gap
possibility on LaFePO and BaFe$_2$(As$_{0.67}$P$_{0.33}$)$_2$.

\end{abstract}

\maketitle

\section{Introduction}
Despite the intensive research effort since the discovery of the
Fe-based superconductors, \cite{Kamihara} the pairing symmetry of
this new class of superconducting (SC) compounds has not yet been
settled. Early theories and experiments appear to best support the
sign changing $s$-wave pairing state (denoted as $s_{\pm}$-,
$s^{\pm}$-, or $\pm s$-state in the literature).
\cite{Mazin,Kuroki,other-theory1,other-theory2,other-theory3}
However, there exist several Fe-pnictide compounds that are not
seemingly compatible with the $\pm$s state but are strongly
suggesting for the presence of nodes in their SC states.
Among others, LaFePO, \cite{Fletcher,Hicks}
BaFe$_2$(As$_{0.67}$P$_{0.33}$)$_2$,\cite{BaFe2AsP2_pene} and
KFe$_2$As$_2$ \cite{KFe2As2_pene} are the most compelling
compounds for the nodal gap, and in lesser degree
Ba(Fe$_{1-x}$Co$_x$)$_2$As$_2$ \cite{Reid} is also suspected.

Commonly taken evidences for the nodal gap in the above mentioned
compounds are: (1) $T$-linear temperature dependence of
penetration depth $\lambda(T)$ down to very low
temperatures,\cite{Fletcher,Hicks,BaFe2AsP2_pene,KFe2As2_pene} and
(2) a strong field dependence in the thermal conductivity slope
$\lim_{T \rightarrow 0} \kappa(H,T)/T$, which is proportional to
in between $\sqrt{H}$ and $H$, accompanied with a substantial
fraction of the residual thermal conductivity $\lim_{T,H
\rightarrow 0} \kappa(H,T)/T =\kappa_{s0}/T$.
\cite{BaFe2AsP2_pene,Matsuda LFPO,KFe2As2_Dong,KFe2As2_Reid} These
features are the well known signatures of the nodal gap
superconductors such as the $d$-wave superconductivity of the
high-$T_c$ cuprates.
And although it was recently shown that the strong field
dependence of the thermal conductivity $\kappa (H, T \rightarrow
0)$ can be equally well explained with the $\pm$$s$-wave state,
\cite{Bang Volovik} the extremely close $T$-linear $\lambda(T)$ is
hard to be reconciled with other than a clean nodal gap
superconductor. Furthermore, the finite value of the residual
thermal conductivity $\kappa_{s0}/T$ measured in all three
compounds \cite{BaFe2AsP2_pene,Matsuda
LFPO,KFe2As2_Dong,KFe2As2_Reid} --it is known that the nodal gap
SC state produces an universal thermal conductivity slope
independent of the amount of impurity concentrations
\cite{universal1,universal2,universal3} -- is another evidence for
a nodal gap state, so it was widely interpreted to support the
presence of nodes in these compounds together with the penetration
depth data.

In this paper, however, we will show that there is a serious and
unreconcilable conflict between the above mentioned two
experimental evidences for the nodal gap. We notice that (1) the
universal value of $\kappa_{S0}/T$ delivers no information about
the dirtiness of the superconducting sample, however, (2) the
normal state value of $\kappa_N /T$ tells us the amount of dirts
in the sample. Then combining the facts (1) and (2), the ratio
$\kappa_S /\kappa_N$, which is usually plotted data in
experiments, is a very good indicator of the dirtiness of the
sample. Inspecting the reported data of thermal conductivities of
the three compounds, we concluded that the measured samples of
LaFePO and BaFe$_2$(As$_{0.67}$P$_{0.33})_2$ should have a large
amount of impurities and hence cannot be compatible of the
$T$-linear $\lambda(T)$ within the nodal gap scenario.
In the case of KFe$_2$As$_2$, there exist two very different
thermal conductivity data by Dong {\it et al.}\cite{KFe2As2_Dong}
and Reid {\it et al.}\cite{KFe2As2_Reid} with different $T_c$
samples, 3K and 3.8K, respectively. Our analysis of the thermal
conductivity data showed that the sample of \cite{KFe2As2_Reid}
($T_c \sim 3.8$K) is cleaner with at least 10 times less impurity
concentration than the sample of Ref.\cite{KFe2As2_Dong} ($T_c
\sim 3$K). Hence, the former sample of KFe$_2$As$_2$ can possibly
be compatible with the $T$-linear $\lambda(T)$ data. We conclude
that KFe$_2$As$_2$ can remain a possible nodal gap superconductor,
but not LaFePO and BaFe$_2$(As$_{0.67}$P$_{0.33})_2$.

\section{Theory}

Assuming the quasiparticle excitation $E({\bf k})=\sqrt{(v_1
k_1)^2+(v_F k_2)^2}$ in the $d$-wave superconductor ($v_F$ Fermi
velocity perpendicular to the Fermi surface(FS), $v_1$ nodal gap
velocity parallel to the FS), the universal thermal (electric)
conductivity in the nodal gap superconductor has been derived as
follows.\cite{universal1,universal2,universal3}

\barr \frac{\kappa_S (T = 0)}{T} & \sim & v_F ^2 \int dk_1 dk_2
\frac{\gamma_s ^2}{[\gamma_s ^2 + (v_1
k_1)^2+(v_F k_2)^2]^2} \nonumber \\
&=& \frac{v_F ^2}{v_F v_1} \frac{\Delta_0}{\sqrt{\gamma_s ^2
+\Delta_0 ^2}}, \earr

\noindent where $\Delta_0$ is the maximum gap value of the
$d$-wave gap $\Delta(\theta)$ and $\gamma_s$ is the impurity
induced damping rate at zero energy in the SC state. As well
known, $\frac{\kappa_S}{T}$ indeed becomes universal, independent
of the impurity concentrations and scattering strength, but only
in the limit of $\Delta_0 \gg \gamma_s$; Eq.(1) clearly shows that
a deviation occurs when $\gamma_s \sim \Delta_0$. The normal state
limit of the above is easy to derive as

\barr \frac{\kappa_N (T = 0)}{T} &\sim& \frac{v_F ^2}{v_F v_1}
\frac{\Delta_0}{\gamma_n}, \earr

\noindent where $\gamma_n$ is the impurity induced damping rate in
the normal state and in general $\gamma_s \neq \gamma_n$ for the
same impurity strength and concentration. Knowing that the normal
state should have no memory of the superconductivity, the above
expression of $\frac{\kappa_N}{T}$ is a disguised form for
convenient comparison with Eq.(1) and $\frac{\Delta_0}{v_1}$
becomes $\pi/4$ or a momentum scale of the FS size. Also we don't
need to know the material specific parameters like $N(0), v_F$,
etc to estimate the absolute magnitude of the thermal
conductivities because for our purpose we only need the ratio

\barr \frac{\kappa_S}{\kappa_N} &=& \frac{\gamma_n}{\sqrt{\gamma_s
^2 +\Delta_0 ^2}} \\
&=& \frac{\Gamma}{\Delta_0}\frac{1}{\sqrt{1+ \Gamma/\Delta_0}}
\approx \frac{\Gamma}{\Delta_0}. \earr

\noindent where $\Gamma=n_{imp}/\pi N(0)$ is the impurity
concentration parameter and in the second line of the above
equations we used the results of $\gamma_s \approx \sqrt{\Delta_0
\Gamma}$ and $\gamma_n = \Gamma$ assuming the unitary impurity
scattering strength. Eq.(4) is the key  result of this paper.

It was nice to observe the universal value of the thermal
conductivity slope $\kappa_S /T$ of Eq.(1) to confirm a nodal
superconductor. On the other hand, it was also a drawback since
the universal thermal conductivity slope doesn't tell us how dirty
or clean the sample is.  However, as shown in Eq.(4), the ratio
$\kappa_S /\kappa_N$ is an excellent indicator of the dirtiness of
the specific SC sample. At this point, we would like to recall the
fact that the typical experimental values of $\kappa_S /\kappa_N$
measured in LaFePO and BaFe$_2$(As$_{0.67}$P$_{0.33}$)$_2$ are
$\frac{\Gamma}{\Delta_0} \approx 0.2-0.4$ which is quite high
level of impurity concentration for a nodal gap superconductor.

\section{Numerical calculations and discussions}

In this section we will show the full numerical calculations of
the field dependence of $\lim_{T\rightarrow 0}\kappa_S (H,T)/T$ as
well as the specific coefficient $\lim_{T\rightarrow 0} C(H, T)/T$
of the canonical $d$-wave gap ($\Delta(\theta)=\Delta_0 \cos{2
\theta}$) state with the various impurity concentrations. Also the
results of the penetration depth $\lambda(T)$ will be shown for
the corresponding impurity concentrations.

To calculate the field dependencies of the thermal conductivity
and specific heat in the mixed state with applied field, we just
need to calculate the position dependent DOS $N(\omega, r)$ in the
presence of vortices. In the semiclassical approximation, the
matrix form of the single-particle Green's function in the SC
state, including Doppler shift of the quasiparticle excitations
$\epsilon (k)$ due to the circulating supercurrent ${\bf v_s
(r)}$, is given by \cite{Bang Volovik,Volovik1,Volovik2}

\beq
\hat{G} ({\bf k, r,\omega,\theta})=\frac{[\omega +  {\bf v_s (r)}
\cdot {\bf k}] \tau_0 + \epsilon (k) \tau_3 + \Delta _{\theta}
\tau_1}{[ \omega + {\bf v_s (r)} \cdot {\bf k}]^2 - \epsilon ^2
(k) - \Delta ^2 _{\theta}}
\eeq

\noindent where $\tau_i$ are Pauli matrices and the supercurrent
velocity around the vortex core ${\bf v_s (r)}$ is $\sim
\frac{1}{m} \frac{1}{r} {\bf \hat{\theta}}$, with ${\bf r}$ the
distance from the vortex core. The position dependent DOS is
calculated as $N (\omega,r)= - \frac{1}{\pi} {\rm Tr Im} \sum_k
\int d\theta~ G_{0} ({\bf k, r,\omega, \theta})$. Finally, the
field dependent quantities are obtained from the areal average DOS
per unit volume as $\bar{N} (\omega,H)=\int_{\xi} ^{R_H} dr^2 N
(\omega,r) / \pi R_H ^2$ with the magnetic length $R_H =
\sqrt{\frac{\Phi_0}{\pi H}}$ ($\Phi_0$ a flux quanta) and the SC
coherence length $\xi$.

The impurity scattering is included by the $\mathcal{T}$-matrix
method. \cite{T-mtx1,T-mtx2,Bang-imp} The impurity induced
self-energies renormalize the frequency and order parameter (OP)
as $\omega \rightarrow \tilde{\omega} =\omega +
\Sigma^{0}(\omega)$ and $\Delta_{0} \rightarrow \tilde{\Delta}_0 =
\Delta_0 + \Sigma^1 (\omega),$ with $\Sigma ^{0,1} (\omega)  =
\Gamma \cdot \mathcal{T}^{0,1} (\omega)$, where
$\mathcal{T}^{0,1}$ are the Pauli matrices $\tau^{0,1}$ components
of the $\mathcal{T}$-matrices in the Nambu space. However,
$\mathcal{T}^{1}$ is identically zero in the $d$-wave state. Then
all impurity effect and the Volovik effect can be incorporated
into the local Green's function Eq.(5) by replacing $\omega$ by
$\tilde{\omega}$.

After calculating the averaged $\bar{N} (\omega,H)$ for all
frequencies, specific heat is calculated as
\beq C(T,H) = \int_0 ^{\infty} d \omega (\frac{\omega}{T})^2
\frac{\bar{N} (\omega, H)}{{\rm cosh}^2 (\frac{\omega}{2 T})} \eeq
Similarly, thermal conductivity is calculated by \cite{Ambegaoka}
\beq \kappa (T,H,r) \propto  v_F ^2 \int d\theta \int_0 ^{\infty}d
\omega (\frac{\omega}{T})^2 \frac{ K (\omega,H,r, \theta)}{{\rm
cosh}^2 (\frac{\omega}{2 T})} \eeq
with

\beq K (\omega,H,r, \theta) = \frac{1} {Im \sqrt{\tilde{ z}^2 -
\Delta ^2 _{\theta}}} \times \Big(1 + \frac{|\tilde{z}|^2 -
|\Delta_{\theta}| ^2}{|\tilde{ z}^2 - \Delta ^2 _{\theta}|} \Big )
\eeq
\noindent where $\tilde{z} =\tilde{\omega} +{\bf v}_s ({\bf r)}
\cdot {\bf k_F} $. And the longitudinal and transversal thermal
conductivities are calculated as $\kappa _{\parallel}(T,H) =
\int_{\xi} ^{R_H} d^2 r \kappa(T,H,r)  / \pi R_H ^2$ and
$\kappa^{-1} _{\perp}(T,H) = \int_{\xi} ^{R_H} d^2 r
\kappa^{-1}(T,H,r) / \pi R_H ^2 $, respectively.

\begin{figure}
\noindent
\includegraphics[width=130mm]{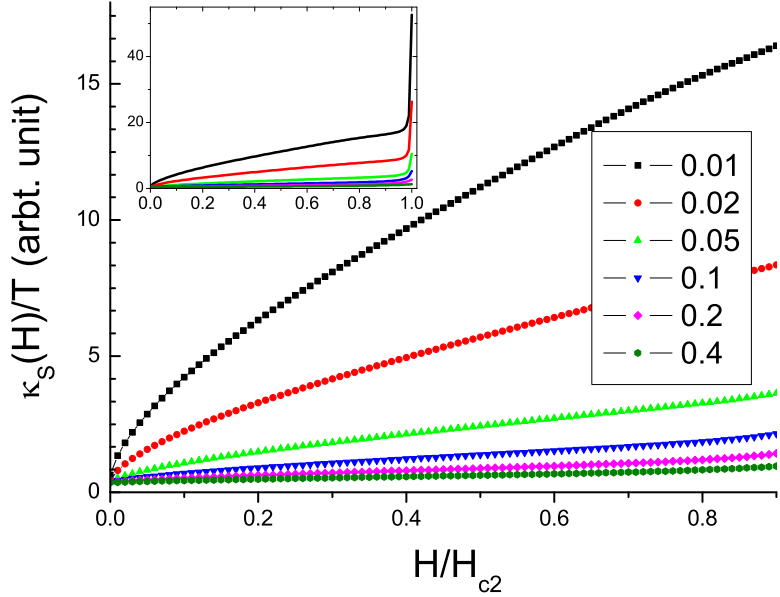}
\caption{(Color online) Thermal conductivity $\kappa_S (H)/T$ vs
the normalized fields $H/H_{c2}$ of the $d$-wave SC state,
calculated at $T=0.02 \Delta_0$ for various impurity
concentrations $\Gamma/\Delta_0= 0.01, 0.02, 0.05, 0.1, 0.2,$ and
0.4. (unitary impurity). The inset shows the full range of fields
up to $H/H_{c2}=1$. \label{fig1}}
\end{figure}

\subsection{Thermal conductivity}
Figure 1 shows the theoretical thermal conductivity
$\kappa(T,H)/T$ vs $H$ of the $d$-wave SC state calculated at the
low temperature limit of $T/\Delta_0 =1/50$ with the varying
impurity concentrations of the unitary scatterers,
$\Gamma/\Delta_0= 0.01, 0.02, 0.05, 0.1, 0.2,$ and 0.4. First, the
results indeed showed that the universal thermal conductivity
$\lim_{T,H \rightarrow 0} \kappa(T,H)/T$ is well reproduced by our
numerical calculations for the vast range of impurity
concentrations. Second, it showed that the normal state limit of
$\kappa (T,H)/T$, which is approached by increasing the field
strength $H$ toward $H_{c2}$, is inversely proportional to the
impurity concentration as shown in Eq.(4). The inset shows the
results for the full range of $H/H_{c2} =[0:1]$ and we can see
that $\kappa(H)/T$ sharply increases near $H_{c2}$. This is due to
a rapid collapse of the gap $\Delta_0 (H)$ towards $H_{c2}$ and
our semiclassical approximation faithfully follows the Doppler
shifting effect of this rapidly collapsing gap up to $H_{c2}$.
While this is the correct calculation results with the
semiclassical approximation, it is also known that this
semiclassical approximation is not precisely correct near $H_{c2}$
where the quantum effect should become important.\cite{BPT} So the
exact field dependence of $\kappa(T,H)/T$ near $H_{c2}$ in Fig.1
should not be taken seriously.

However, the important points for our purpose are: (1) at both
limits, the universal limit value of $\lim_{T,H \rightarrow 0}
\kappa(T,H)/T$ and the normal state limit value
$\kappa(H=H_{c2})/T$ are exact, and (2) the overall field
dependence of the initially slow rise and then a rapid rise of
$\kappa(H)/T$ near $H_{c2}$ is the genuinely correct behavior
regardless of different theoretical treatments. \cite{Mishra} The
main conclusions of this paper relies only on these two points.
The main panel in Fig.1 shows the results for the limited region
of $H \leq 0.9 H_{c2}$ for a better resolution of the low field
behavior of $\kappa_S (H)/T$.

\begin{figure}
\noindent
\includegraphics[width=130mm]{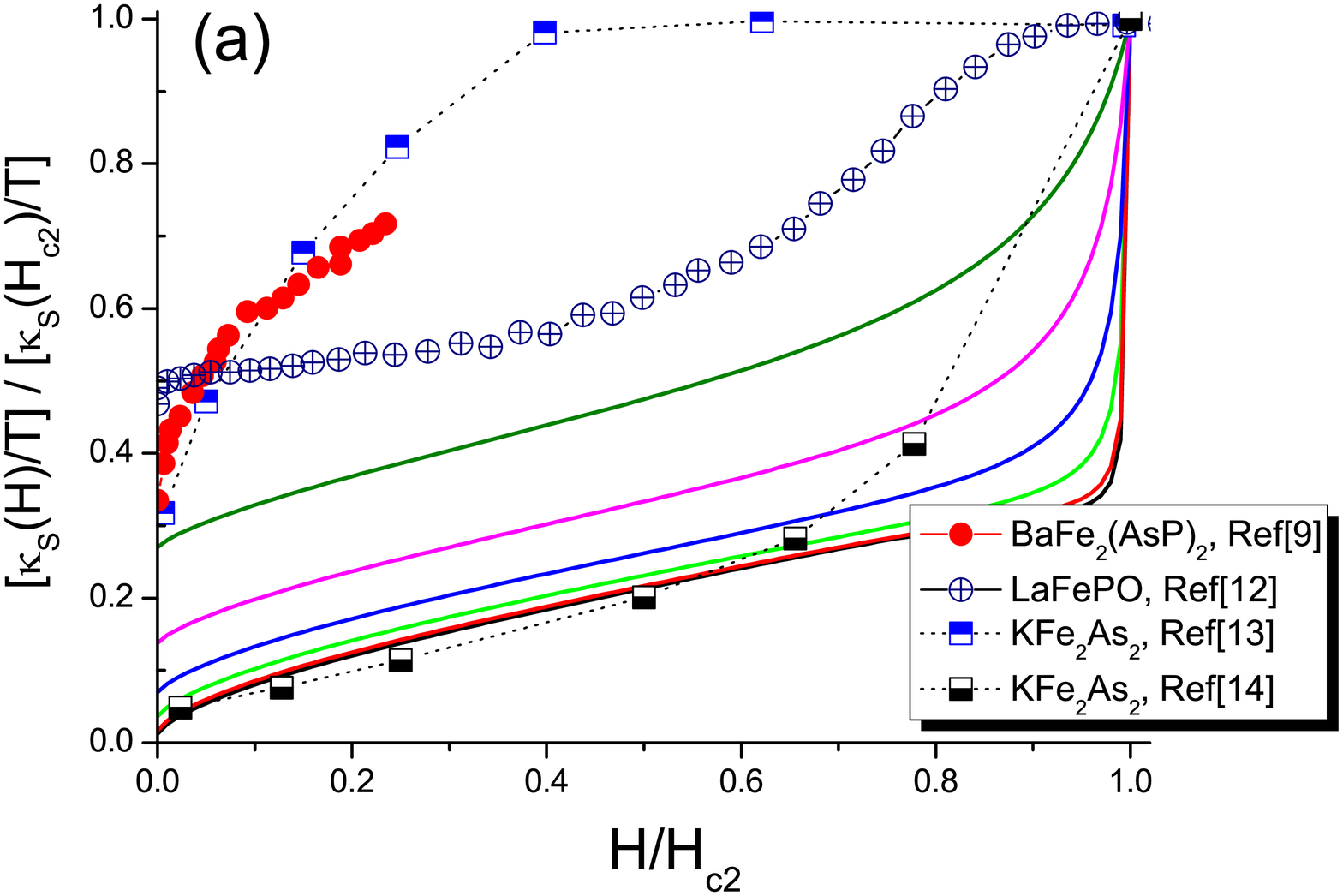} \\
\includegraphics[width=130mm]{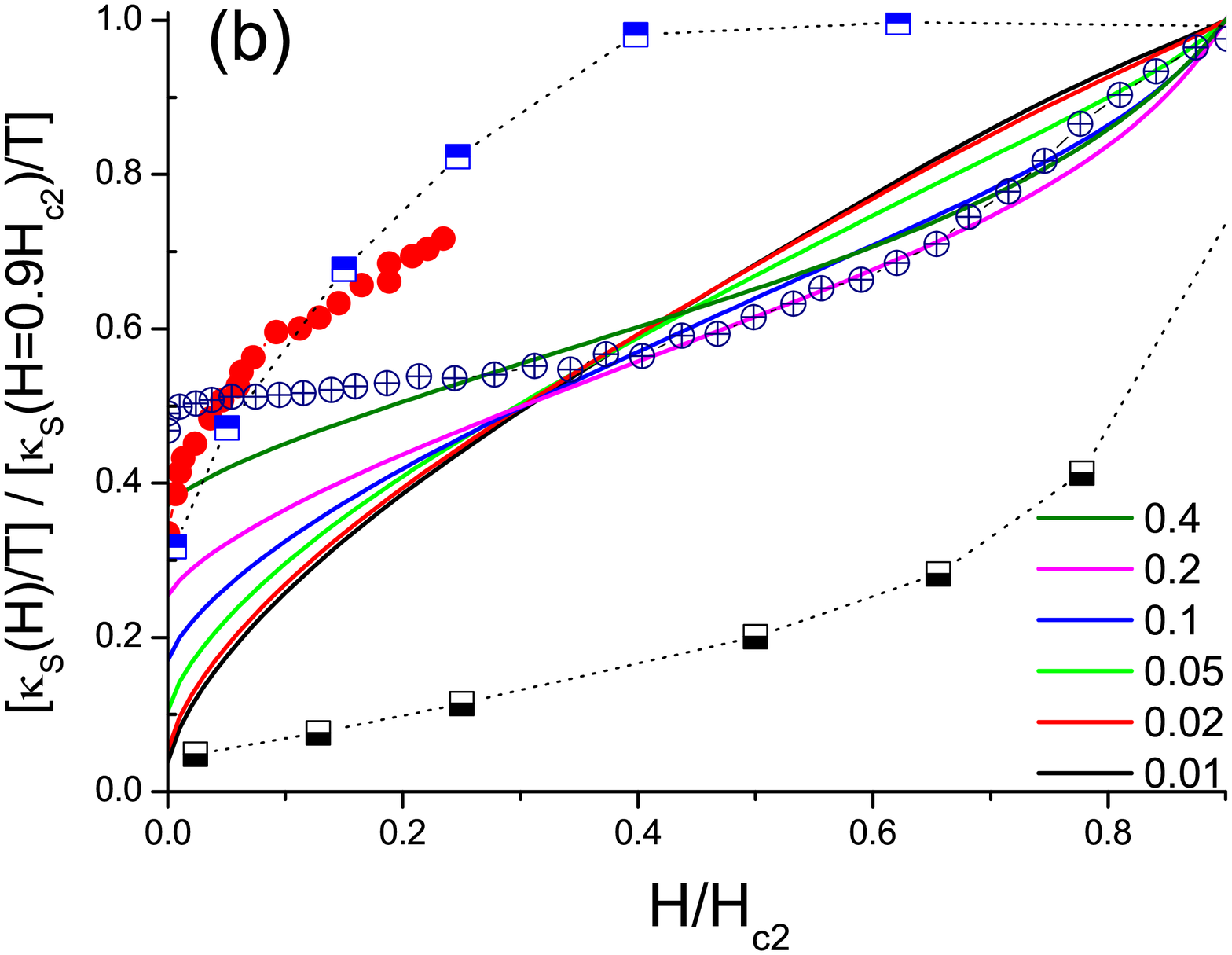}
\caption{(Color online) (a) Theoretical thermal conductivities
(lines) of Fig.1 normalized by the values of $\kappa_S (H_{c2})/T$
for each impurity concentrations $\Gamma/\Delta_0= 0.01, 0.02,
0.05, 0.1, 0.2,$ and 0.4. (the same colors as in Fig.1). Symbols
are the cropped data of experiments.
(b) The same data as in (a) but normalized by the values of
$\kappa_S (H=0.9 H_{c2})/T$ for each impurity concentration.
Symbols are the same as in (a). \label{fig2}}
\end{figure}

In Fig.2, we replotted the theoretical results of Fig.1 with two
different normalizations and the experimental data of
$[\kappa_S(H)/T] / [\kappa_N/T]$ are overlaid. Fig.2(a) normalized
$\kappa_S(H)/T$ of Fig.1 with its $\kappa_S(H=H_{c2})/T$ values
for each impurity concentration, and Fig.2(b) used the
$\kappa_S(H=0.9 H_{c2})/T$ values for normalization. The second
normalization plot by the $H=0.9 H_{c2}$ values was chosen because
it is the typical point of saturation before the sharp rise as
seen in the inset of Fig.1. and this concave-down saturation
behavior approaching $H_{c2}$ is the typical observations in
experiments. \cite{BaFe2AsP2_pene,Matsuda LFPO,KFe2As2_Dong}
Different normalizations yield different line shapes of the
normalized $\kappa_S(H)/T$ which is supposed to be compared to the
experimental $[\kappa_S(H)/T] / [\kappa_N/T]$. The true behavior
should be somewhere in between Fig.2(a) and Fig.2(b), but we
emphasize that this fine detail is irrelevant to our main
conclusions and analysis. The overlayed experimental data are
BaFe$_2$(As$_{0.67}$P$_{0.33}$)$_2$, \cite{BaFe2AsP2_pene} LaFePO,
\cite{Matsuda LFPO} and KFe$_2$As$_2$. \cite{KFe2As2_Dong,
KFe2As2_Reid}

Regardless of the choice of the normalizations, the experimental
values of the residual thermal conductivity of
BaFe$_2$(As$_{0.67}$P$_{0.33}$)$_2$, \cite{BaFe2AsP2_pene} LaFePO,
\cite{Matsuda LFPO} and KFe$_2$As$_2$ \cite{KFe2As2_Dong}
unambiguously indicate that these compounds should have the
impurity concentration $\Gamma/\Delta_0
> 0.4$, which is extremely dirty superconductor. Our theoretical
calculations are with a single $d$-wave gap band. In reality, if
there exists a nodal gap in these multiband Fe-pnictide compounds,
the total gap function should consist of a nodal gap $+$ one or
two $s$-wave gaps, for example, a nodal $\pm s$-wave
gap.\cite{Matsuda LFPO} If that is the case, the total $\kappa_N
/T$ should increase due to the addition contributions from other
bands. However, these additional $s$-wave gap bands have
negligible contributions to the residual thermal conductivity
$\lim_{T,H \rightarrow 0}\kappa_S(H)/T $ because they are fully
gapped at low fields and low temperatures. Therefore, we need to
have even higher impurity concentration than $\Gamma/\Delta_0 >
0.4$ in order to match the experimental data\cite{BaFe2AsP2_pene,
Matsuda LFPO, KFe2As2_Dong} of the normalized residual thermal
conductivity $\lim_{H,T \rightarrow 0} [\kappa_S(H)/T] /
[\kappa_N/T]$.

On the other hand, the data of KFe$_2$As$_2$ by Reid {\it et al.}
\cite{KFe2As2_Reid} is very different from the data of
KFe$_2$As$_2$ by Dong {\it et al.} \cite{KFe2As2_Dong} We can see
that the data of Reid {\it et al.} \cite{KFe2As2_Reid} reasonably
fit the theoretical result in Fig.2(a) with the impurity
concentration, $\Gamma/\Delta_0 < 0.02$, which is relatively clean
limit. As discussed in Ref.\cite{KFe2As2_Reid}, the discrepancy
between the data of two groups is understood by the sample purity.
Judging from the $T_c$ of two samples (3.80K and 3K, respectively)
and our theoretical calculations of $\kappa_S/\kappa_N$ in Fig.2,
the sample of Reid {\it et al.} must be much cleaner, by about
10-20 times, than the one of Dong {\it et al.} and it appears to
be consistent with the result of the clean d-wave calculation with
$\Gamma/\Delta_0 < 0.02$ in Fig.2(a).

Summarizing the cases of BaFe$_2$(As$_{0.67}$P$_{0.33}$)$_2$
\cite{BaFe2AsP2_pene} and LaFePO \cite{Matsuda LFPO}, if we
interpret the thermal conductivity data of these two compounds
with a nodal gap scenario, we are led to conclude that both
compounds are very dirty nodal gap superconductors. However, if
this is true, the linear-$T$ penetration depth
measurements\cite{Fletcher,Hicks,BaFe2AsP2_pene} of these two
compounds are in a serious conflict with the dirty nodal gap
scenario.

In the case of KFe$_2$As$_2$, we have two seemingly contradicting
thermal conductivity measurements \cite{KFe2As2_Dong,
KFe2As2_Reid} as seen in Fig.2(a) and (b). However, despite the
large difference of $T_c$ and the line shapes of
$\kappa_S(H)/\kappa_N$, both samples reported a similar value of
the residual thermal conductivity: $\kappa_{S0}/T$=3.7 $\pm$ 0.4
mW/K$^2$cm (Ref.\cite{KFe2As2_Reid}) and $\kappa_{S0}/T$=2.27
$\pm$ 0.02 mW/K$^2$cm (Ref.\cite{KFe2As2_Dong}, here we ignored
the correction by geometric factor discussed in
Ref.\cite{KFe2As2_Reid}), respectively. This fact itself is a
strong supporting evidence for the nodal gap in KFe$_2$As$_2$
compound. The different zero field intercepts of the data of two
samples in Fig.2 are due to the normalization by the normal state
thermal conductivities: $\kappa_N /T$=7.36 $\pm$ 0.04 mW/K$^2$cm
for Dong {\it et al.} \cite{KFe2As2_Dong} and $\kappa_N /T
\approx$ 109 mW/K$^2$cm for Reid {\it et al.} \cite{KFe2As2_Reid}
The data of Reid {\it et al.}\cite{KFe2As2_Reid} can fit
reasonably well with the clean nodal gap calculation
($\Gamma/\Delta_0 <0.02$) for the most of low field region as seen
in Fig.2(a). The data of Dong {\it et al.},\cite{KFe2As2_Dong}
however, doesn't fit with any calculational results in Fig.2.
While the residual thermal conductivity value of it can be fit
with a dirty nodal gap with $\Gamma/\Delta_0 \approx 0.4$, the
data for $H >0$ increases much rapidly at low fields and saturates
to become flat for $H > 0.4 H_{c2}$, not even close to any
theoretical results in Fig.2. But, if we assume the additional
bands with $s$-wave gaps in addition to a nodal gap, as in a nodal
$\pm s$-wave state, it would be possible to fit the data of both
clean and dirty limits of Ref.\cite{KFe2As2_Dong,KFe2As2_Reid}.
The detailed SC properties of the nodal $\pm s$-wave state will be
reported in future publication.

\begin{figure}
\noindent
\includegraphics[width=130mm]{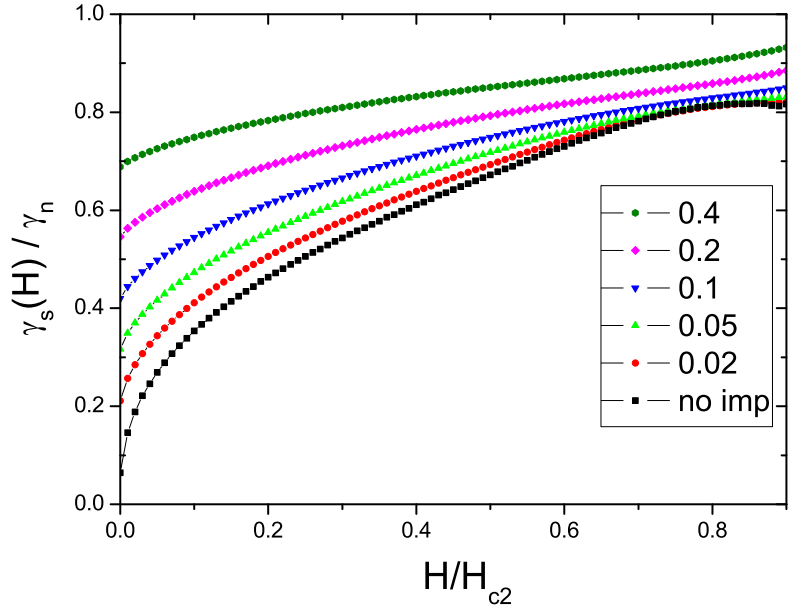}
\caption{(Color online) Normalized specific heat coefficient
$\lim_{T \rightarrow 0} C(T,H)/T=\gamma_S (H)$ vs fields
$H/H_{c2}$ calculated at $T=0.02 \Delta_0$ of the $d$-wave SC
state for various impurity concentrations $\Gamma/\Delta_0= 0,
0.02, 0.05, 0.1, 0.2,$ and 0.4 (unitary impurity). \label{fig3}}
\end{figure}

\begin{figure}
\noindent
\includegraphics[width=130mm]{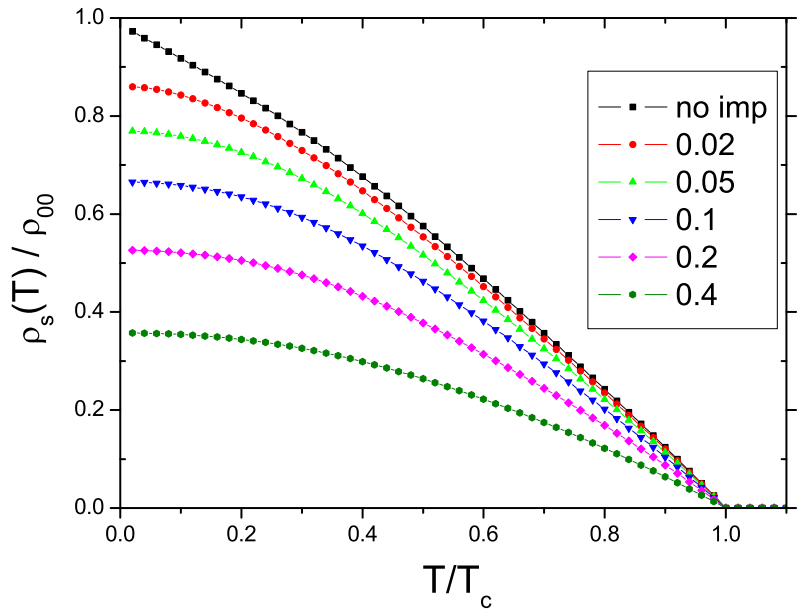}
\caption{(Color online) Normalized superfluid density $\rho_S(T)$
vs $T/T_c$ of the $d$-wave SC state for various impurity
concentrations $\Gamma/\Delta_0= 0, 0.02, 0.05, 0.1, 0.2,$ and 0.4
(unitary impurity). $2 \Delta_0 /T_c =4$ is used.\label{fig4}}
\end{figure}

\subsection{Specific heat coefficient and Superfluid density}

To foster the above discussions, we calculated the field
dependence of the specific heat coefficient $\gamma_s(H)=\lim_{T
\rightarrow 0} C(H,T)/T$ and the temperature dependence of the
superfluid density $\rho(T) \sim 1/\lambda^2(T)$ of the $d$-wave
state with various concentrations of the unitary impurities as in
Figs.1 and 2. Figure 3 is the normalized $\gamma_s(H)/\gamma_n$
for $\Gamma/\Delta_0 =0.4, 0.2, 0.1, 0.05, 0.02,$ and 0 (no
impurity limit). It shows the expected behavior as the $\sqrt{H}$
behavior of $\gamma_s(H)$ in the clean limit becomes flattened
with increasing the impurity concentration. The only point that we
want to emphasize for our purpose is that even a small amount of
impurities, for example, $\Gamma/\Delta_0 =0.02$, immediately
creates a substantial fraction of the specific heat coefficient
$\gamma_s(H=0) \approx 0.2 \gamma_n$. This demonstrates that a
nodal gap such as the $d$-wave state is extremely vulnerable to
the unitary impurity scattering to create the low energy
excitations.

Figure 4 shows the normalized superfluid density $\rho_S(T) \sim
1/\lambda^2(T)$ of the $d$-wave state with corresponding impurity
concentrations of Fig.3. It also shows the well known behavior of
$\rho_S(T)$ of the $d$-wave state with impurities. The typical
$T$-linear $\rho_S(T)$ behavior in the clean limit changes to the
$T^2$ behavior at low temperatures with impurities. Similarly to
the evolution of the specific heat coefficients in Fig.3, even a
small amount of impurities changes quite a wide temperature region
into the $T^2$ behavior. For example, the impurity concentration
of $\Gamma/\Delta_0 =0.02$ makes $\rho_S(T) \sim T^2$ for $0 < T <
0.2T_c$. In view of the fact that all three pnictide compounds
studied in this paper reported the $T$-linear behaviors of
$\lambda(T)$ down to extremely low temperatures: LaFePO ( $0.02<
T/T_c$, \cite{Fletcher} and $ 0.08 < T/T_c$ \cite{Hicks}),
BaFe$_2$(As$_{0.67}$P$_{0.33}$)$_2$  ($0.025 < T/T_c$
\cite{BaFe2AsP2_pene}), and KFe$_2$As$_2$ ($0.05 < T/T_c$
\cite{KFe2As2_pene}), the required purity of these samples for the
nodal gap scenario is $\Gamma/\Delta_0 < 0.01$ or even cleaner.
However, this clean nodal gap scenario is totally contradicting to
the thermal conductivity measurements $\kappa_S (H)/T$ in the
cases of BaFe$_2$(As$_{0.67}$P$_{0.33}$)$_2$ \cite{BaFe2AsP2_pene}
and LaFePO, \cite{Matsuda LFPO} but possibly not with the case of
KFe$_2$As$_2$.\cite{KFe2As2_Dong, KFe2As2_Reid}

\section{Remark on the quantum oscillation (QO) experiments} It
is well known that the observation of the QO such as de Haas-van
Alphen (dHvA) oscillation is possible only with very clean samples
and, therefore, often quoted as an indication of the extreme
purity of the probed sample. All three compounds studied in this
paper have reported the QO experiments \cite{LFPO_QO, KFeAs_QO,
BaFeAsP_QO}, hence we should worry about the consistency between
the QO experiments -- which warrant that the probed samples are
clean -- and our analysis of the thermal conductivity -- which
indicate that most of theses compounds are not so clean. To begin
with, we note that the criteria of the cleanness for the QO signal
and the SC properties are different; the former is $\Gamma < \hbar
\omega_{c} (=\hbar eB/m^{*}; m^{*}$=renormalized mass) and the
latter is $\Gamma < \Delta_0$. Also, we need some interpretation
for the typical dirtiness deduced from our thermal conductivity
analysis, i.e., $\Gamma / \Delta_0$, which was based on the single
$d$-wave model. We concluded in the previous sections that only
KFe$_2$As$_2$ is possibly consistent with a nodal gap, but LaFePO
and BaFe$_2$(As$_{0.67}$P$_{0.33}$)$_2$ are not consistent with a
nodal gap state but would be more consistent with a $\pm s$-wave
gap model with a small isotropic gap on the major band and a
larger isotropic gap on the minor band, namely, $\Delta_S \ll
\Delta_L$ and $N(0)_S \gg N(0)_L$. Therefore, in the cases of
LaFePO and BaFe$_2$(As$_{0.67}$P$_{0.33}$)$_2$, for example, the
deduced damping rate $\Gamma / \Delta_0 \sim 0.4$ should be
understood as $\Gamma / \Delta_S \sim 0.4$ while $T_c$ of the
compound is mostly governed by $\Delta_L$.\cite{Bang_model}
Bearing these in mind, let us
examine the cases of each compound below.\\

\subsection {LaFePO ($T_c \simeq 6K$).} The QO measurement on LaFePO has
observed signals with magnetic fields above $\sim 9$T but the
practically useful signals were obtained above $\sim 20$T and up
to 45T.\cite{LFPO_QO} The estimated cyclotron frequency at $B=20$T
is $\hbar \omega_{c} \sim 1.2 - 2.4 meV$  with the renormalized
mass $m^{*}/m_0$ ($m_0$, free electron mass) $\sim 1 -
2$.\cite{LFPO_QO}  and the estimated damping rate from our
analysis is $\Gamma=0.4 \Delta_0 \sim 0.35 meV$ assuming the BCS
relation $\Delta_0/T_c =1.75$. Therefore, the condition for the QO
observation $\hbar \omega_{c} > \Gamma$ sufficiently holds for all
fields $B > 20$T, and, therefore, without invoking further
argument of the multiple gaps, $\Delta_S$ and $\Delta_L$, the
observation of the QO in LaFePO has no contradiction with the
damping rate estimated in our analysis.

\subsection {KFe$_2$As$_2$ ($T_c \simeq$ 3K).} The QO signals on
KFe$_2$As$_2$ \cite{KFeAs_QO} were obtained in the field range of
10 to 17.5 T. The estimated cyclotron frequency is $\hbar
\omega_{c} \sim 0.2 meV$ at $B=10$T with the heavily renormalized
mass $m^{*}/m_0 \sim 6$ of this compound.\cite{KFeAs_QO} As
discussed in the previous section, there exist two very different
thermal conductivity experiments\cite{KFe2As2_Dong, KFe2As2_Reid};
the sample by Reid {\it et al.}\cite{KFe2As2_Reid} seems to be
clean ($\Gamma < 0.02 \Delta_0$) and the other one by Dong {\it et
al.}\cite{KFe2As2_Dong} seems to be dirtier ($\Gamma \approx 0.4
\Delta_0)$.
The estimated damping rates are $\Gamma \sim 0.0087 meV$ for the
clean one and $\Gamma \sim 0.175 meV$ for the dirty one,
respectively, assuming the BCS relation $\Delta_0/T_c =1.75$ with
$T_c =$3K. Therefore, if the sample used for the QO
experiment\cite{KFeAs_QO} is close to the cleaner one, there is
absolutely no problem to observe the QO signals ($\hbar \omega_{c}
\gg \Gamma$; $\hbar \omega_{c} \sim 0.2 meV$, $\Gamma \sim 0.0087
meV$). On the other hand, if the sample were on the side of the
dirtier one, the observation of the QO signals should be very weak
at best ($\Gamma \sim 0.175 meV, \hbar \omega_{c} \sim 0.2 meV$).
Putting together, there exists a wide range of sample purity
between $0.0087 meV < \Gamma < 0.175 meV$ with which the QO
experiment was possible.

\subsection{BaFe$_2$(As$_{0.67}$P$_{0.33}$)$_2$ ($T_c \simeq 30$K).}

Shishido {\it et al.}\cite{BaFeAsP_QO} have performed the QO
experiments with BaFe$_2$(As$_{1-x}$P$_{x}$)$_2$. However, the QO
signals was obtained only with $1 \geq x \geq 0.41$ for the field
range from 17T to 55T, and the $x=0.33$ sample never produced
meaningful signals up to 55T. On top of that, even in the samples
of $1 \geq x \geq 0.41$ only the electron band FSs ($\alpha$ and
$\beta$ bands in their notations) produced signals but the hole
band FSs never produced measurable signals. With theses, we can
estimate the overall damping rate of the $x=0.33$ sample should be
higher than $\hbar \omega_{c} \sim 0.55 meV$ using the
renormalized mass $m^{*}/m_0 \approx 3$ and the maximum field
strength $B=55$T used in experiments.\cite{BaFeAsP_QO} On the
other hand, our estimated damping rate is $\Gamma=0.4 \Delta_0
\sim 1.4 meV$ using the BCS relation and $T_c$=30K. So it is
consistent with the failure of the QO experiment for the $x=0.33$
sample. In reality, since we have argued that the $\pm s$-wave
state is more consistent with BaFe$_2$(As$_{0.67}$P$_{0.33}$)$_2$,
if we understood $\Gamma=0.4 \Delta_0$ as $\Gamma=0.4 \Delta_S$,
the real damping rate $\Gamma$ should be $< 1.4 meV$ but still $>
0.55 meV$.

\section{Conclusions}
In conclusion, we have carefully reexamined the experimental
evidences for the possible existence of the SC gap nodes in the
three most suspected Fe-pnictide compounds, LaFePO,
BaFe$_2$(As$_{0.67}$P$_{0.33}$)$_2$, and KFe$_2$As$_2$. We have
derived an exact relation for the ratio between the universal
residual thermal conductivity $\kappa_S/T$ and its normal state
value $\kappa_N /T$ in the $d$-wave state. Using this ratio
$\kappa_S/\kappa_N \approx \Gamma/\Delta_0$ as an indicator to
determine the dirtiness of the SC sample, we have shown that the
reported experimental data of the thermal conductivity in
BaFe$_2$(As$_{0.67}$P$_{0.33}$)$_2$ \cite{BaFe2AsP2_pene} and
LaFePO \cite{Matsuda LFPO} indicated that the measured samples are
the dirty limit superconductors, hence contradicting to the clean
limit nodal gap scenario deduced from the penetration depth
measurements. \cite{Fletcher, Hicks, BaFe2AsP2_pene} To this end,
if the nodal gap scenario fails for these two compounds, we
propose a dirty $\pm s$-wave gap state as a possible scenario to
reconcile the apparently contradicting experiments of thermal
conductivity and penetration depth measurements -- the large
residual thermal conductivity slope $\kappa_S /T$ and the
$T$-linear $\lambda(T)$; in this scenario one isotropic $s$-wave
gap is much smaller than the other one and the small gap is almost
filled with the impurity band caused by a sufficient amount of
impurity scattering.

In the case of KFe$_2$As$_2$, there exist two qualitatively
different data of the thermal conductivity measurements.
\cite{KFe2As2_Dong, KFe2As2_Reid} It appears that the one of Reid
{\it et al.}\cite{KFe2As2_Reid} is a clean sample but the one by
Dong {\it et al.}\cite{KFe2As2_Dong} contains at least 10 times
more impurities. We concluded that the clean sample result can be
consistent with a nodal gap scenario both for the thermal
conductivity and penetration depth measurements. On the other
hand, the thermal conductivity data of the dirty
sample\cite{KFe2As2_Dong} can be understood with a dirty nodal gap
plus additional isotropic gaps as in the nodal $\pm s$-wave gap,
but the $T$-linear $\lambda(T)$ cannot be compatible with this
sample by any scenario. Therefore, the gap symmetry of
KFe$_2$As$_2$ should be further investigated by the
cross-examination of the penetration depth and thermal
conductivity measurements with samples with various purities.

\ack{This work was supported by Grants No. NRF-2010-0009523 and
No. NRF-2011-0017079 funded by the National Research Foundation of
Korea.}

\end{document}